\documentclass[conference]{IEEEtran}
\IEEEoverridecommandlockouts
\usepackage{multirow} 
\usepackage{hyperref} 
\usepackage{url}      
\usepackage{caption}
\usepackage[numbers,sort&compress]{natbib}

\usepackage{amsmath,amssymb,amsfonts}
\usepackage{algorithmic}
\usepackage{graphicx}
\usepackage{textcomp}
\usepackage{xcolor}
\def\BibTeX{{\rm B\kern-.05em{\sc i\kern-.025em b}\kern-.08em
    T\kern-.1667em\lower.7ex\hbox{E}\kern-.125emX}}
\begin{document}

\title{AudioEditor: A Training-Free Diffusion-Based Audio Editing Framework\\
\thanks{$^{*}$ Corresponding author.}
}

\author{\IEEEauthorblockN{Yuhang Jia$^{1}$, Yang Chen$^{1}$, Jinghua Zhao$^{1}$, Shiwan Zhao$^{1}$, Wenjia Zeng$^{2}$, Yong Chen$^{2}$, and Yong Qin$^{1,*}$}
\IEEEauthorblockA{$^1$\textit{College of Computer Science, Nankai University}, Tianjin, China, \\
$^2$\textit{Lingxi (Beijing) Technology Co., Ltd.}, Beijing, China\\
Email: 2013628@mail.nankai.edu.cn,
qinyong@nankai.edu.cn}
}

\maketitle

\begin{abstract}
Diffusion-based text-to-audio (TTA) generation has made substantial progress, leveraging latent diffusion model (LDM) to produce high-quality, diverse and instruction-relevant audios. However, beyond generation, the task of audio editing remains equally important but has received comparatively little attention. Audio editing tasks face two primary challenges: executing precise edits and preserving the unedited sections. While workflows based on LDMs have effectively addressed these challenges in the field of image processing, similar approaches have been scarcely applied to audio editing. In this paper, we introduce AudioEditor, a training-free audio editing framework built on the pretrained diffusion-based TTA model. AudioEditor incorporates Null-text Inversion and EOT-suppression methods, enabling the model to preserve original audio features while executing accurate edits. Comprehensive objective and subjective experiments validate the effectiveness of AudioEditor in delivering high-quality audio edits. Code and demo can be found at \href{https://github.com/NKU-HLT/AudioEditor}{https://github.com/NKU-HLT/AudioEditor}.
\end{abstract}

\section{Introduction}
\vspace{-3pt}
Diffusion-based algorithms \cite{ho2020denoising,song2020denoising} have recently gained widespread use in generative tasks for images, speech, and audio. It is stable to train and can produce high-quality, diverse results based on the given text prompt \cite{rombach2022high}. Following the success of diffusion-based text-to-image (TTI) generation \cite{rombach2022high,ramesh2021zero,saharia2022photorealistic,zhou2024storydiffusion}, significant progress has also been made in text-to-audio (TTA) generation \cite{liu2023audioldm,liu2024audioldm,huang2023make,huang2023make2,ghosal2023text,majumder2024tango}. These TTA systems have leveraged the LDM architecture and algorithms, generating high-quality and high instruction-relevance audios. A recent work, Auffusion \cite{xue2024auffusion}, fine-tuned the pre-trained TTI model with a large amount of audio dataset, resulting in significant improvements and high-performance in audio generation quality and text-audio alignment.

While significant advancements have been made in audio generation, the ability to precisely edit existing audio has received comparatively little attention. In image processing filed, many works have leveraged the generative capabilities of diffusion model to invert real data back into the model's domain \cite{song2020denoising,dhariwal2021diffusion} and then reconstruct it, resulting in a variety of image editing workflows. SDEdit \cite{meng2021sdedit} edits images by adding noise to a real image and then denoising it with a new description using LDM. P2P \cite{hertz2022prompt} achieves image adjustment and editing by modifying the attention maps between text and image embeddings during the execution of diffusion and denoising. Work\cite{huberman2024edit} recently proposed a specific noise space for DDPM inversion, which has been shown to be edit-friendly in image processing. Another recent work \cite{li2024get} guides diffusion models to generate edited images that align with target prompts by adjusting the prompt embedding based on its editable and preserved components.

As for audio editing, AUDIT \cite{wang2023audit} is the first known audio editing model, which trained a dedicated LDM under supervised learning by building a triplet training dataset, consisting of input audio, instruction and output audio. While providing a specialized solution, it is costly in terms of data resources and depends on specific-format editing instructions(e.g., “add...at...”), which are inflexible. WavCraft \cite{liang2024wavcraft} leverages large language models to interpret editing instructions and performs audio editing by invoking external audio and speech models. Recent works \cite{manor2024zeroshot} and PPAE \cite{xu2024promptguided} have utilized image processing techniques, such as DDPM inversion and attention map manipulation, for audio editing, demonstrating the feasibility and effectiveness of performing audio editing with diffusion-inversion methods initially.

In this paper, we propose AudioEditor, an audio editing framework based on \cite{li2024get} with the following key advantages:  1) \textbf{Flexible Editing}: No specific-format instructions are needed—users simply provide a target caption and specify the text edit region, and AudioEditor automatically locates and edits the corresponding audio components. 2) \textbf{Competitive Performance}: We significantly enhance two particularly challenging aspects of audio editing—executing precise edits and preserving original audio features—by innovatively incorporating techniques from image processing, such as Null-text Inversion and EOT-suppression. 3) \textbf{Training-Free}: Leveraging only a pre-trained diffusion-based TTA model, our framework delivers high-quality audio editing without requiring training or fine-tuning on specific editing datasets.

\section{METHOD}

\begin{figure*}[t]
\centerline{\includegraphics[width=1.0\textwidth]{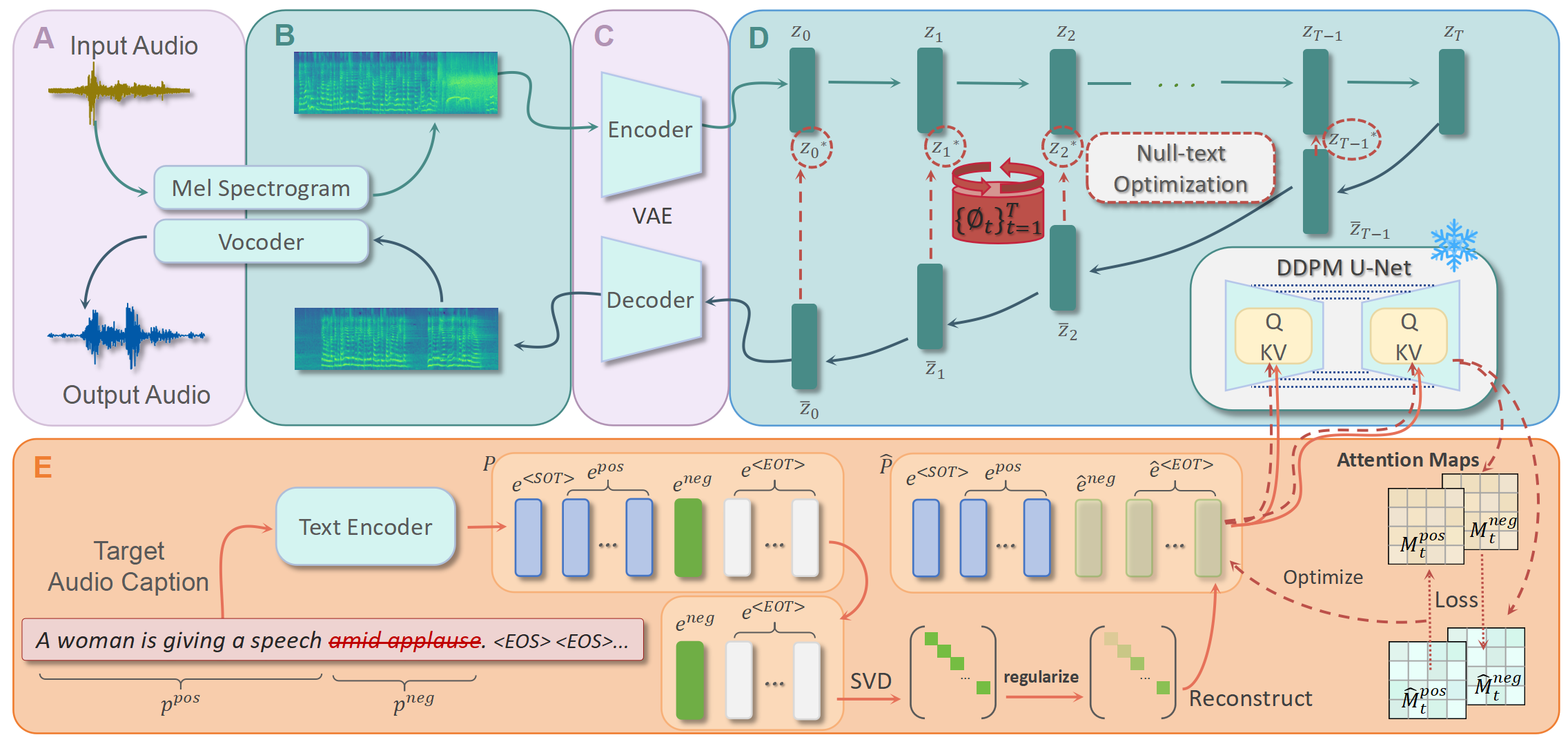}}
\caption{The overall workflow of AudioEditor. The workflow can be divided into 5 parts: A) audio space processing, B) spectrogram space processing, C) latent space processing, D) performing DDIM Inversion and Null-text Optimization, and E) performing EOT-suppression and Attention Loss updating.} 
\label{fig1}
\end{figure*}
\vspace{-3pt}
In this work we utilize TTA model Auffusion as foundational framework for AudioEditor. Fig.~\ref{fig1} illustrates the complete pipeline of AudioEditor performing audio editing. Given an audio clip $\mathcal{X}$ and a target prompt $p$ as description, we first convert $\mathcal{X}$ to mel-spectrogram and use a VAE to transform the mel-spectrogram into latent representation ${z}_{0}$. Then we apply T-steps DDIM inversion and Null-text Optimization to obtain lantent representation ${z}_{T}$ and the optimized null-text prompts ${\{\varnothing_{t}\}_{t=1}^{T}}$, which will be used in reconstruction process.

For text prompt input $p$, we first use the text encoder of Auffusion to obtain the prompt embedding $\mathcal{P} = \{e^{<SOT>}, e^{p}_{0}, ...,e^{p}_{|p|-1}, e^{<EOT>}, ..., e^{<EOT>} \}$. We then apply EOT-suppression method to remove the parts of $\mathcal{P}$ that require editing while preserving the rest, obtaining $\hat{\mathcal{P}}$. Subsequently, the processed $\hat{\mathcal{P}}$ will be used as condition for denoising and reconstruction process in LDM. During the denoising process, we introduce the attention loss $\mathcal{L}_{A}$ to further enhance the effectiveness of EOT-suppression.
\vspace{-2pt}
\subsection{Diffusion based Text-to-Audio Model}\label{AA}
To accomplish audio editing tasks with latent diffusion, a pretrained diffusion-based TTA model is required. Here we have Auffusion \cite{xue2024auffusion}, which has been fine-tuned on a large amount of audio datasets based on pretrained TTI diffusion model. Auffusion comprises four main components: 1) a Variational Auto-Encoder (VAE) to convert inputs between spectral space and latent space; 2) a latent diffusion model for diffusion and denoising; 3) a CLIP-based text encoder $\psi$ to embed input text; and 4) a HiFi-Gan as vocoder.

Similar to many contemporary latent diffusion models, we employ the classifier-free guidance technique as described in \cite{ho2022classifier} to mitigate the enhanced effect triggered by conditioned text. Specifically, if we denote $\mathcal{P} = \psi(\textit{prompt})$ as the embedding of prompt and $\varnothing = \psi(\textit{`'})$ as the embedding of null text, then one-step classifier-free guidance prediction can be defined as follows:
\begin{equation}
\tilde{\varepsilon}_{\theta}\left(z_{t}, t, \mathcal{P}, \varnothing\right)=w \cdot \varepsilon_{\theta}\left(z_{t}, t, \mathcal{P}\right)+(1-w) \cdot \varepsilon_{\theta}\left(z_{t}, t, \varnothing\right).
\end{equation}
, where $w$ is the guidance scale parameter, which defaults to 7.5 or 1.0 in AudioEditor.

\subsection{DDIM Inversion and Null-text Optimization}\label{AA}
It is believed that the ODE process is reversible within the limit of small steps \cite{mokady2023null,song2020denoising,dhariwal2021diffusion}. Based on this assumption, we can perform the denoising process of DDIM in reverse direction from $z_{0}$ to $z_{T}$. The DDIM Inversion with $\varnothing = \psi(\textit{`'})$ can be defined as follows:
\begin{equation}
z_{t+1}=\sqrt{\frac{\alpha_{t+1}}{\alpha_{t}}} z_{t}+\left(\sqrt{\frac{1-\alpha_{t+1}}{\alpha_{t+1}}}-\sqrt{\frac{1-\alpha_{t}}{\alpha_{t}}}\right) \cdot \varepsilon_{\theta}\left(z_{t}, t, \mathcal{P}, \varnothing\right) .
\end{equation}

Using DDIM Inversion, we can invert a real data input from $z_{0}$ to $z_{T}$, and then reconstruct it from $z_{T}$ to $\bar{z_{0}}$. However, due to the domain discrepancy between pretrained model and real input data, the reconstructed audio often exhibits significant differences from the original. To address this issue, we introduce Null-text Optimization \cite{mokady2023null}, which modifies only the constant $\varnothing$ in classifier-free guidance inference without altering the model weights or the inference pipeline.

Considering that with different $w$ values, $\varnothing$ influences the inference to varying extents, Null-text Optimization first sets $w=1$ to perform DDIM inversion on the input $z_{0}$, resulting in an intermediate latent trajectory $\{z_{t}^{*}\}_{t=0}^{T}$. Next, with $w=7.5$, we initialize $\bar{z_{T}}=z_{T}^{*}$, $\varnothing_{T} = \psi(\textit{`'})$, and perform diffusion process from $\bar{z_{T}}$ to $\bar{z_{0}}$. At each diffusion step $t$ , $\varnothing_{t} = \psi(\textit{`'})$ is optimized at a rate of $\eta$, guiding the denoising latent state $\bar{z_{t}}$ towards the reversion latent state $z_{t}^{*}$, aiming for high-fidelity reconstruction:
\begin{equation}
\varnothing_{t} \leftarrow \varnothing_{t}-\eta \nabla_{\varnothing}\left\|z_{t-1}^{*}-z_{t-1}\left(z_{t}, t, \mathcal{P}, \varnothing_{t}\right)\right\|_{2}^{2}
\end{equation}
At the end of each optimization step, we updated $z_{t-1}$ and $\varnothing_{t}$ as follows:
\begin{equation}
\bar{z}_{t-1} = z_{t-1}\left(z_{t}, t, \mathcal{P}, \varnothing_{t}\right), \ \varnothing_{t-1} = \varnothing_{t}
\end{equation}

In AudioEditor, we iterate 10 times per step $t$ to optimize $\varnothing_{t}$. After optimization, we get T optimized null text embeddings ${\{\varnothing_{t}\}_{t=1}^{T}}$, which will replace the global constant $\varnothing = \psi(\textit{`'})$ in the subsequent reconstruction process. With the help of Null-text Optimization, we achieve high-fidelity reconstruction of audio, making audio editing with LDMs possible.
\subsection{EOT-suppression}\label{AA}
Our experiments have demonstrated that even with DDIM Inversion and Null-text Optimization, directly employing the embedding of target prompts as condition will still lead to mismatching with intended edits. Inspired by the work \cite{li2024get}, which analyzed the information carried by End Of Tokens (EOT) embedding and successfully guided image synthesis and editing through the EOT-suppression method. We decided to incorporate EOT-suppression into our AudioEditor.

We first categorize the elements in prompt embedding $\mathcal{P} = \{e^{<SOT>}, e^{p}_{0}, ...,e^{p}_{|p|-1}, e^{<EOT>}, ..., e^{<EOT>} \}$ into two types: negative and positive. Negative embedding tokens correspond to the parts of target prompt that need to be edited, including those to be deleted, added, or replaced. Positive embedding tokens represent the remaining parts. Thus, we obtain ${\mathcal{P}^{\prime}} = \{e^{<SOT>}, e^{pos}, ...,e^{neg}, e^{<EOT>}, ..., e^{<EOT>} \}$. We then extract the negative embedding and EOT embedding, resulting in ${\mathcal{X}} = \{e^{neg}, e^{<EOT>}, ..., e^{<EOT>} \}$. We apply Singular Value Decomposition (SVD) to ${\mathcal{X}}$: ${\mathcal{X}} = \boldsymbol{U} \boldsymbol{\Sigma} \boldsymbol{V}^{T}$, where $\boldsymbol{\Sigma} = \operatorname{diag}\left(\sigma_{0}, \sigma_{1}, \cdots, \sigma_{n}\right)$, $\sigma_{0} \geq \sigma_{1} \geq ... \geq \sigma_{n}$.

According to the observations in \cite{li2024get} and \cite{gu2014weighted} on image processing tasks, noise primarily resides in the bottom-K singular values, while the negative parts are mainly found in the top-K singular values. Therefore, for deletion operations, the top-K singular values should be suppressed, whereas for replacement or addition operations, they should be enhanced. Based on this insight, We adopt the approach from \cite{li2024get} to regularize $\boldsymbol{\Sigma}$ to $\widehat{\boldsymbol{\Sigma}}$ using the following rule:
\begin{equation}
\hat{\sigma}=\beta \cdot e^{\alpha \sigma} * \sigma.
\end{equation}
, where $\beta = 1.0$, $\alpha = 1.0$ for deletion, while $\beta = 1.2$, $\alpha = 0.001$ for addition and replacement.

After above regularization, we reconstruct ${\widehat{\mathcal{X}}}$: ${\widehat{\mathcal{X}}} = \boldsymbol{U} \widehat{\boldsymbol{\Sigma}} \boldsymbol{V}^{T}$ and then $\widehat{\mathcal{P}}$, which will be used as the condition for denoising in AudioEditor.
\subsection{Attention Loss}\label{AA}
To further preserve the positive parts and suppress the negative parts in prompt embedding, we utilize the method proposed in \cite{hertz2022prompt} to extract the cross-attention maps between the regularized prompt embedding $\widehat{\mathcal{P}}$ and audio embedding, obtaining$\{\widehat{\mathcal{A}}^{pos}_{t}, \widehat{\mathcal{A}}^{neg}_{t}\}$ at each step $t$. Similarly, we extract cross-attention maps for ${\mathcal{P}}$ to obtain $\{{\mathcal{A}}^{pos}_{t}, {\mathcal{A}}^{neg}_{t}\}$. ${\mathcal{A}}^{pos}$ and ${\mathcal{A}}^{neg}$ are corresponding to the attention maps of the positive and negative parts respectively. 

Then we introduce the following attention loss, building on the approach in \cite{li2024get}, to further update prompt embedding $\widehat{\mathcal{P}}$ at each step:
\begin{equation}
\mathcal{L}_{\mathcal{A}}=\lambda_{pos} \left\|\widehat{\mathcal{A}}_{t}^{pos}-\mathcal{A}_{t}^{pos}\right\|^{2} - \lambda_{neg} \left\|\widehat{\mathcal{A}}_{t}^{neg}-\mathcal{A}_{t}^{neg}\right\|^{2}
\end{equation}
, where $\lambda_{pos} = 1.0$ and $\lambda_{neg} = 0.5$ in AudioEditor.

\section{EXPERIMENTS AND RESULTS}

\subsection{Dataset}
Considering that Auffusion was fine-tuned with datasets such as AudioCaps \cite{kim2019audiocaps}, WavCaps \cite{mei2024wavcaps}, and MACS \cite{martin2021ground}, we choose the AudioCaps \textit{testset} as our evaluation dataset. We manually select 300 audio samples from this set, selecting ones that are particularly suitable for editing. These 300 samples are then categorized into three groups—addition, deletion, and replacement on specific principles: audio containing a single event is deemed suitable for addition, while audio with multiple events is appropriate for replacement or deletion. Finally, we manually craft target caption for each audio sample based on its content and original caption to guide editing. Both the test audio samples and target captions are open-sourced in our code repository to facilitate the reproduction of our experiment results.
\subsection{Baselines and Experimental Setup} 
To establish a baseline, we adopt a similar approach to AUDIT by using image editing models trained on audio datasets. Specifically, AUDIT uses SDEdit as its baseline, and we do the same. Inspired by \cite{molad2023dreamix}, we also use TTA model Auffusion to directly generate target audio based on the target prompt, which serves as a benchmark for assessing the model’s editing capabilities. Additionally, we use the original audio as another benchmark to evaluate the model’s ability to preserve original audio features.
\subsection{Evaluation Metrics}
\textbf{Objective Metrics.} For objective evaluation, we use a pre-trained CLAP model to compute the cosine similarity between the target text prompt and the edited audio, which serves as an indicator of the quality of the audio edits. To measure the similarity between two audio samples or two sets of audio samples, we utilize metrics including Frechet Distance (FD), Frechet Audio Distance (FAD), and Kullback–Leibler (KL) divergence. For assessing the overall audio quality, we use the Inception Score (IS). The FD, KL, and IS metrics are calculated using the PANNs \cite{kong2020panns} classifier, while the FAD is computed with the VGGish \cite{hershey2017cnn} model. The tools for calculating FD, FAD, KL, and IS are referenced from project\footnote{\hypertarget{link1}\url{https://github.com/haoheliu/audioldm\_eval}}, and the pre-trained CLAP model is sourced from project\footnote{\hypertarget{link1}\url{https://github.com/microsoft/CLAP}}.

\textbf{Subjective Metrics.} For subjective evaluation, we use Mean Opinion Score (MOS) \cite{loizou2011speech} to assess the overall quality of an audio, which we refer to as \textit{Quality}. To evaluate the model's ability to preserve the original audio features, we use Similarity MOS (SMOS) to measure the similarity between the edited and original audios, which we refer to as \textit{Faithfulness}. Additionally, we ask listeners to judge the relevance between the edited audio and the target caption, to assess the quality of the edits, which we call \textit{Relevance}.
% \vspace{-5pt}

\subsection{Main Results and Analysis}

\textbf{Objective Evaluation.} As shown in Table \ref{tab1}, our model outperforms both the SDEdit and Original\_wavs under the CLAP metric, coming very close to the audio directly regenerated from target description. This demonstrates our model’s strong editing capability. The IS score indicates that our editing approach does not significantly degrade the overall quality of the original audio.

\begin{table}[h!]
\begin{center}
\renewcommand{\arraystretch}{1.2} 
\captionsetup{font=small} 
\caption{Objective Evaluation Results(a).} 
\begin{tabular}{ c|c c}
\hline
\textbf{Editing\_}&\multicolumn{2}{|c }{\textbf{Objective Metrics}} \\
% \cline{2-4} 
\textbf{Models} & \hspace{3mm} \textbf{\textit{Clap}}$\uparrow$ & 
\textbf{\textit{IS}}$\uparrow$ \\
\hline
Original\_wavs&  \hspace{3mm} 48.2\% &  4.77\\
Regenerated\_wavs&  \hspace{3mm} \textbf{59.2\%}& 5.19 \\
SDEdit(baseline)& \hspace{3mm} 56.8\%& 5.55 \\
AudioEditor(ours)& \hspace{3mm} \hspace{1mm} 57.6\% $^{\mathrm{*}}$& 5.19  \\
\hline
\multicolumn{3}{l}{$^{\mathrm{*}}$ denotes a suboptimal value, which, in certain metrics, may} \\
\multicolumn{3}{l}{\hspace{2mm} indicates a more desirable result than the optimal one.} 
\end{tabular}
\label{tab1}
\end{center}
\end{table}
\vspace{-10pt}
In the similarity test with Regenerated\_wavs (Table \ref{tab2}), our model largely surpasses both SDEdit and Original\_wavs, achieving a closer match to the audio regenerated from target caption, further highlighting its superior editing performance.

\begin{table}[h!]
\begin{center}
\renewcommand{\arraystretch}{1.2} 
\captionsetup{font=small} 
\caption{Objective Evaluation Results(b).} 
\begin{tabular}{ c|c c c }
\hline
\textbf{Editing\_}&\multicolumn{3}{|c }{\textbf{Similarity with (Regenerated\_wavs)}} \\
% \cline{2-4} 
\textbf{Models} &\hspace{2mm} \textbf{\textit{FD}}$\downarrow$ \hspace{3mm}& \hspace{3mm} \textbf{\textit{FAD}}$\downarrow$& \textbf{\textit{KL}}$\downarrow$ \\
\hline
Original\_wavs& 55.48& \hspace{3mm} 6.45& 2.73 \\
SDEdit(baseline)& \textbf{37.59} & \hspace{3mm} 4.21 & 1.12 \\
% \hline
AudioEditor(ours)& 37.63 &\hspace{3mm} \textbf{3.27} & \textbf{1.07} \\
\hline
\end{tabular}
\label{tab2}
\end{center}
\end{table}
\vspace{-10pt}
Additionally, in the similarity test with Original\_wavs (Table \ref{tab3}), our model is significantly closer to the original audio compared to Regenerated\_wavs, while being marginally better than SDEdit. This indicates that our model effectively preserves the characteristics of the original audio while performing high-quality edits.

\textbf{Subjective Evaluation.} From Table \ref{tab4}, we observe that in terms of overall quality (Quality), our model effectively preserves audio quality better than both the regenerated audio and the baseline. In the 
\begin{table}[h!]
\begin{center}
\renewcommand{\arraystretch}{1.2} 
\captionsetup{font=small} 
\caption{Objective Evaluation Results(c).}
\begin{tabular}{ c|c c c }
\hline
\textbf{Editing\_}&\multicolumn{3}{|c }{\textbf{Similarity with (Original\_wavs)}} \\
% \cline{2-4} 
\textbf{Models} &\hspace{2mm} \textbf{\textit{FD}}$\downarrow$ \hspace{3mm}& \hspace{2mm}\textbf{\textit{FAD}}$\downarrow$& \textbf{\textit{KL}}$\downarrow$ \\
\hline
Regenerated\_wavs& 55.48& 6.45& 2.61 \\
SDEdit(baseline)& 44.13 & 5.68 & \textbf{1.79} \\
% \hline
AudioEditor(ours)& \textbf{43.48}  &\textbf{4.95} & 1.93 \\
\hline
\end{tabular}
\label{tab3}
\end{center}
\end{table}
% \vspace{-8pt}
editing quality (Relevance) metric, our model's relevance to the target description is very close to that of the regenerated audio and shows significant improvement over the original audio and the baseline, demonstrating its high editing quality. Regarding the preservation of original features (Faithfulness), our model exhibits a significant improvement 
in similarity to the original audio compared to both the regenerated audio and the baseline, proving its effectiveness in retaining the original audio characteristics.
\begin{table}[h!]
\begin{center}
\renewcommand{\arraystretch}{1.2} 
\captionsetup{font=small} 
\caption{Subjective Evaluation Results.} 
\begin{tabular}{ c|c c c }
\hline
\textbf{Editing\_}&\multicolumn{3}{|c }{\textbf{Subjective Metrics}} \\
% \cline{2-4} 
\textbf{Models} &\textbf{\textit{Quality}}$\uparrow$\hspace{3mm}& \textbf{\textit{Relevance}}$\uparrow$& \textbf{\textit{Faithfulness}}$\uparrow$ \\
\hline
Original\_wavs& \textbf{4.22}& 1.54& \textbf{5.00} \\
Regenerated\_wavs& 4.11& \textbf{3.92}& 2.03 \\
SDEdit(baseline)& 4.11& 3.14& 3.77 \\
% \hline
AudioEditor(ours)&\hspace{1mm} 4.13 $^{\mathrm{*}}$ &\hspace{1mm} 3.88 $^{\mathrm{*}}$ & \hspace{1mm}3.82$^{\mathrm{*}}$ \\
\hline
\end{tabular}
\label{tab4}
\end{center}
\end{table}
\vspace{-10pt}

\subsection{Ablation Study}
\begin{table}[h!]
\begin{center}
\renewcommand{\arraystretch}{1.2} 
\captionsetup{font=small} 
\caption{Ablation Study Results.} 
\begin{tabular}{ c|c c c c}
\hline
\textbf{Editing\_}&\multicolumn{4}{|c }{\textbf{Objective Metrics}} \\
% \cline{2-4} 
\textbf{Models} &\textbf{\textit{Clap}}$\uparrow$\hspace{3mm}& \textbf{\textit{FD(ori)}}$\downarrow$ &
\textbf{\textit{FAD(ori)}}$\downarrow$ &
\textbf{\textit{IS}}$\uparrow$ \\
\hline
AudioEditor(ours)& 57.6\% & 43.38 & 4.95& 5.19 \\
w/o Null-opt& 57.2\%& \hspace{1mm} 45.54 $\uparrow$ & \hspace{1mm} 5.33 $\uparrow$ & 5.35 \\
w/o EOT-sup& \hspace{1mm} 56.5\% $\downarrow$ & 42.78& 4.91& 5.39 \\
% \hline
\hline
\end{tabular}
\label{tab5}
\end{center}
\end{table}
\vspace{-10pt}
As shown in Table \ref{tab5}, after removing Null-text Optimization, the model's FAD and FD metrics show a significant negative impact, indicating a reduced ability to preserve the original audio features. When EOT-suppression method is removed, the model’s CLAP score drops, reflecting a decrease in editing capability. In terms of the IS metric, our two approaches only have a minor impact on the overall audio quality, without causing any significant deterioration. These ablation results align with our expectations that Null-text Optimization enhances feature retention, while EOT-suppression improves editing precision. This demonstrates the effectiveness and necessity of each method within our framework.

\section*{CONCLUSION}

In this paper, we proposed AudioEditor, a training-free audio editing framework built upon the pre-trained diffusion-based TTA model. AudioEditor harnesses the invertible nature of latent diffusion and incorporates advanced image processing techniques, such as Null-Text Inversion and EOT-suppression, overcoming the challenges of executing precise editing and preserving the unedited sections. The effectiveness of AudioEditor is demonstrated through its competitive performance in both objective and subjective experiments, establishing it as one of a pioneering framework for diffusion-based audio editing.

\small
\bibliographystyle{ieeetr}
\bibliography{ref}

\end{document}